\documentclass[fleqn,10pt,twoside]{article}

\usepackage[english] {babel}

\usepackage {graphicx}
\usepackage {graphics}
\usepackage {floatflt}
\usepackage {latexsym}
\usepackage {caption3} 
\usepackage[labelsep=period]{caption}[2007/11/04 v.3.1e]
\usepackage {amssymb}

\def\noi{\noindent}

\renewcommand{\thesubsubsection}%
        {\arabic{section}.\arabic{subsection}.\arabic{subsubsection}.}

\newcommand{\heads}[2]{\markboth{\protect\small\it #1}{\protect\small\it #2}}

\newcommand{\Arthead}[5]{ \setcounter{page}{#4}\thispagestyle{empty}\noi
    \unitlength=1pt \begin{picture}(500,40)

        \put(0,58){\shortstack[l]{\small\it ISSN 0202-2893, Gravitation \& Cosmology,
                          #1, Vol.#2, No. #3, pp. #4--#5    
\footnotesize\copyright \ Pleiades Publishing Ltd., 2011.} }

    \end{picture}
	 }     		
\def\prepno#1#2
    {\thispagestyle{empty}
    \noi    \unitlength=1mm
    	\begin{picture}(178,10)
            \put(177,15){\llap{\bf #1}}
            \put(177,10){\llap{\small\rm #2}}
        \end{picture}
          }             

\newcommand{\Title}[1]{\noi {\uppercase{\Large #1}}     }
\newcommand{\Author}[2]{\noi{\large\bf #1}\\[2ex]\noindent{\it #2}   }

\newcommand{\Abstract}[1]{\vskip 2mm \begin{center}
        \parbox{16.4cm}{\small\noi #1} \end{center}\medskip}

\newcommand{\foom}[1]{\protect\footnotemark[#1]}

\newcommand{\email}[2]{\footnotetext[#1]{e-mail: #2}
		\addtocounter{footnote}{1}}

\heads{A.M.Baranov}
      {THE RELATIVISTIC FLUID BALL AS A STRATIFIED MODEL OF AN ASTROPHYSICAL OBJECT}

\begin{document}
\twocolumn 
[
\Arthead{2011}{17}{2}{173}{175.}

\vspace{0.5cm}
\begin{center}
\Title{The Relativistic Fluid Ball as a Stratified Model  

\vspace{0.2cm}
   of an Astrophysical Object\foom3}
\end{center}

\begin{center}
\vspace{.5cm}
   \Author{A.M.Baranov\foom1 and A.Yu.Osipov\foom2}   
{\it Dep. of Theoretical Physics, Siberian Federal University,
Svobodny prosp. 79, Krasnoyarsk 660041, Russia}

{\it Received October 3, 2010}
\end{center}

\Abstract
{A relativistic fluid ball with an inhomogeneous static stratified matter configuration is considered. A model of an astrophysical object with this structure of matter is constructed.}

{\bf DOI:} 10.1134/S020228931102006x
\vspace{1.0cm}
] 

\email 1 {alex\_m\_bar@mail.ru; AMBaranov@sfu-kras.ru}
\email 2 {ya.oayu@mail.ru}
\footnotetext[3]{Talk given at the International Conference RUDN-10, 
 June 28--July 3, 2010, PFUR, Moscow.}

\section{Introduction}

\indent

The existense of a huge gradient of the gravitational field in massive astrophysical objects leads to radial separation of matter . In other words, in a strong gravitational field there are strata with the different equations of state. Each strata has its equation of state. In particular a neutron star has a structure of this kind.

A relativistic fluid ball with an inhomogeneous static stratified matter configuration is considered. The ball is filled with a Pascal perfect fluid. The distribution of matter in the star is supposed to be spherically symmetric. The processes connected with rotation and radiation are excluded. The interior model with a stratified structure of mass density is constructed. 

We use the geometric system of units with the velocity of light  $c=1$ and the Newton gravitational constant $G_N = 1,$ i.e. $\varkappa =8\pi.$ The metric is taken in Bondi's form
$$
ds^2 = G(r)^2 dt^2+2L(r)dtdr
$$
$$
-r^2({d\theta}^2+{\sin{\theta }}^2 {d\varphi}^2), 
\eqno{(1)}
$$
where $G(r)^2$ and $L(r)$ are functions from the radial variable $r;$ $t$ is the time coordinate; 
$\theta$ and $\varphi$ are angle variables. Further we will denote $d/dr$ as a prime.

\section{The Einstein equations}

The gravitational field, described by the metric $g_{ik},$ can be found from the Einstein equations 
$$
R_{ik} - \frac{1}{2}Rg_{ik} = - \varkappa \cdot T_{ik},
\eqno{(1)}
$$
where $R_{ik}$ is the Ricci tensor and $R$ is the scalar curvature.

The energy-momentum tensor of a Pascal perfect fluid can be written as 
$$
T_{ik} = \left( {p\left( r \right) + \mu \left( r \right)} \right) \cdot u_i u_k  - p\left( r \right)g_{ik},
\eqno{(3)}
$$
where $p = p(r)$ is the pressure, $\mu = \mu(r)$ is the density of matter, and $u_i$ is the 4-velocity. 

The gravitational equations in dimensionless variables can be written in the form convenient for numerical calculations after elementary transformations: 
$$
\varepsilon (x) = 1 - \frac{\chi }{x} \cdot \int {\mu (x) \cdot x^2 dx};
\eqno{(4)}
$$
$$
G^{\prime\prime}+ \left( {\frac{{\varepsilon '}}{{2 \cdot \varepsilon }} - \frac{1}{x}} \right) \cdot G^\prime + \left( {\frac{{\varepsilon '}}{{2 \cdot x \cdot \varepsilon }} + \frac{{1 - \varepsilon }}{{x^2  \cdot \varepsilon }}} \right) \cdot G = 0;
\eqno{(5)}
$$
$$
p' = - \frac{1}{{2 \cdot \varepsilon }} \cdot \left( {\chi \cdot x \cdot p + \frac{{1 - \varepsilon }}{x}} \right) \cdot \left( {\mu  + p} \right),
\eqno{(6)}
$$
where $x = r/R$ is the dimensionless radius; differentiation in $x$ is denoted by a prime; 
$R$ is the radius of the astrophysical object; $\chi=\varkappa\cdot R^2;$ 
$$
\varepsilon (x) = {G ^2(x)}/{L ^2(x)}.
\eqno{(7)}
$$

In our case, the following matching conditions at $x=1$ can be written
$$
\varepsilon (x=1) = 1- \eta;\;\;G (x=1) = \sqrt {1- \eta};
\eqno{(8)}
$$

$$
G' (x=1) = \frac{\eta}{2 \cdot \sqrt{1- \eta}};
\eqno{(9)}
$$
$$
p (x=1) = 0,
\eqno{(10)}
$$
where $\eta = 2 m/{R}$ is the compactness factor, and $m$ is the stellar mass.

\section{Stratified structure}

The four functions $F(x)$, $L(x)$, $p(x)$, and $\mu(x)$ are unknowns, and we must specify one of them. For example, we take $\mu(x).$ 

We will add the mass density profile to Einstein's equations. This function can be constructed as a multi-step continuous function. The mass density distribution is simulated here by a function of this kind ([1], [2])
$$
 \mu (x) = \mu _0 \cdot \left( {1 - b \cdot x + \displaystyle\frac{{\sin (a \cdot x)^2 }}{a}} \right),
\eqno{(1)}
$$
where $\mu _0$ is the central density, $a$ is a dimensionless parameter ($0 < a < \infty$) which controls the number of layers (see Fig.1), $b$ is a dimensionless parameter connected with the stellar properties.


\begin{figure}[t,h]
\center{\includegraphics[width=0.95 \linewidth]{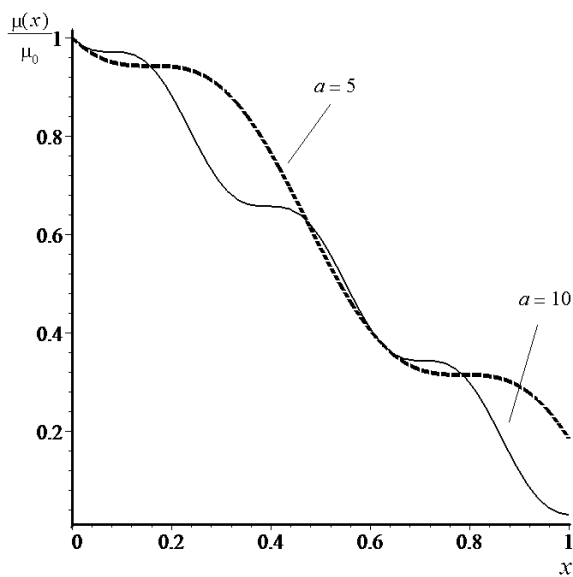}}
\caption{The mass density distribution. There are two strata for $a=5$ and three strata 
for $a=10$, the parameter $b=1$ is fixed.}
\end{figure}



The functions $G(x)$, $L(x)$, $p(x)$, $K(x)=p(x)/{\mu(x)}$ are found numerically by the Runge-Kutta-Fehlberg method. Here we must require the validity of the dominant energy condition, $|p (x)| < \mu(x)$, $p > 0$, $\mu  > 0.$ Matching to the Schwarzschild exterior solution is carried out at $\eta = 0.1$. 

These functions can be plotted for different values of the parameter $a$ with fixed values of $b$. 
For example, we present a plot of $p (x)$ (see Fig.2). 
%

\begin{figure}[t,h]
\center{\includegraphics[width=0.95 \linewidth]{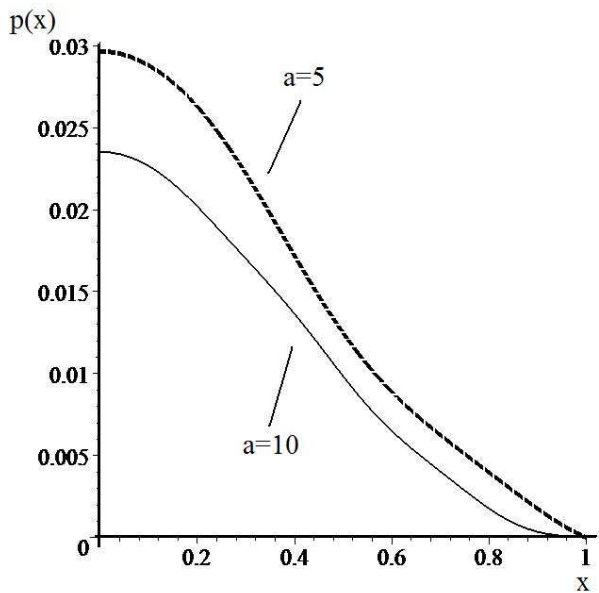}}
\caption{The pressure. The parameter values are $a=5$, $a=10$, $b=1$.}
\end{figure}



\section{Neutron star}

Now we will consider a neutron star model and find the compactness $\eta$ for such an object. 

We have the following equation for parameters $a$, $b$ from the condition (8)
$$
\int_0^1 {\mu \left( x \right)x^2 dx = \frac{{2m}}{{\chi R}}}.
\eqno{(12)}
$$

Its solution depends on the stellar mass $m$ and radius $R,$ and the central density $\rho_0$. This mass 
density is connected with the energy density $\mu_0$ by $\mu_0 = \rho_0 \cdot c^2,$ where $c$ is the speed of light; here and henceforth we work in the CGS units.

We take the characteristic physical parameters of a neutron star: 
$$
R = 1.5 \cdot 10^6 cm,\;\;m = 1.9891 \cdot 10^{33} g,
\eqno{(13)}
$$

$$
\rho _0 = 4.8064 \cdot 10^{14} g/cm^3.
\eqno{(14)}
$$

Then the solution of Eq.(12) gives $a = 10$, for $b = 1$.

The star is filled with neutrons. Accordingly we will use the parametric equation of state of degenerate fermi-gas:
$$
\rho_0  = \left( {\sinh \left( \xi  \right) - \xi } \right)K/c^2;
$$
$$
p_0 = \left( {\sinh \left( \xi  \right) - 8\sinh \left( {\xi /2} \right) + 3\xi } \right)K/3,
\eqno{(15)}
$$
where $K = {{m_f^4 c^5 } \mathord{\left/
 {\vphantom {{m_f^4 c^5 } {\left( {32\pi ^2 \hbar ^3 } \right)}}} \right.
 \kern-\nulldelimiterspace} {\left( {32\pi ^2 \hbar ^3 } \right)}}$, $m_f$ is a rest mass of fermion, $\hbar$ is the Plank constant, and $\xi$ is a parameter, $0 < \xi  < \infty $.

Let us find the mass density $\mu_0$ and the pressure $p_0$ at the stellar center from this equation. 
We have for $\xi = 1$
$$
\rho _0=1.0038\cdot 10^{14} g/cm^3,
$$
$$
p_0=1.07\cdot 10^{33} dyn/cm^2.
\eqno{(16)}
$$

The compactness factor is found from Eq.(12) for the parameter values $a = 10$ and $\mu_0 = 1.0038 \cdot 10^{14} \cdot c^2.$ As a result, $\eta \approx 0.16.$ On the other hand, the characteristic compactness factor for neutron stars found from the observations is $\eta \approx 0.1.$

\section{Gravitational phase 
transitions}

\indent

Besides, we must note that such a model with stratified structure has the Petrov algebraic type changing from layer to layer. In other words, an extremum point exists in each layer, with the first and the second derivatives of the mass density are equal to zero. In particular, such a point is the centre of the ball under study. In neighborhoods of such points, the metric function $g_{00}$ will coincide with that in the interior Schwarzschild solution with a high precision, with its own integration constants. It means that the gravitational field belongs to the algebraic type {\it\bfseries 0} (a conformally flat type). In the other points of the model, the gravitational field belongs to the algebraic type {\it\bfseries D}. Thus in the interior part of our model the algebraic type of the gravitational field varies: into the centre (and its neighborhood) is the type {\it\bfseries 0} which passes on to type {\it\bfseries D}. 

From the point of view of phase transition theory, we have a second-order phase transition. The phases of a "substance" here are the algebraic types of gravitational fields [3]. On the other hand, we have an cusp catastrophe from the point of view of the catastrophe theory. Further on, in the the second layer there is also a second-order phase transition in the gravitational field from type {\it\bfseries D} to type {\it\bfseries 0} and again to type {\it\bfseries D},and so on up to the ball surface.The Schwarzschid exterior solution belongs to the type {\it\bfseries D} as we know. 

In other words, as we move from the center to the surface, an alternation of the algebraic types of gravitational fields by the second-order phase transitions is observed.

Thus the properties of the model confirm the theorem from [4] which says that any spherical gravitational field can only belong to two algebraic types: {\it\bfseries D} and {\it\bfseries 0}. 

Another example of phase transitions of the gravitational fields as transitions from one algebraic type to another can be found in [5].

\section{Summary}

In conclusion, we must note that the model of an astrophysical object (a neutron star ) with a stratified structure has been constructed as a result of studing a relativistic liquid ball. This model allows for obtaining a qualitative description of real stars with stratified structure. Furthermore, in such stars we can expect the existence of the gravitational phase transitions as the alternation of Petrov's algebraic types.


\small
\begin{thebibliography}{99}

\bibitem{1}
 A.M.Baranov and A.Yu.Osipov, {\it Transactions of NKSF-XXXVIII } ( Krasnoyarsk, SibFU, 2009 ), ~ p.122 ~ (in Russian).
\bibitem{2}
 A.M.Baranov and A.Yu.Osipov,  {\it Transactions $\;$ of ~ Russian school-seminar GRACOS-2010}, p.141 (2010) (in Russian).
\bibitem{3}
 A.M.Baranov, {\it Abstr.of Confer. "200 Years of Kazan University"}, p.107 (2004).
\bibitem{4}
 A.M.Baranov,{\it Vestnik of Krasnoyarsk State University}, Phys. Math. Sci. No.1, 5 (2006).
\bibitem{5}
A.M.Baranov, Zh. SibFU. Mat. Phys. {\bf 4}(1), 3 (2011). 
\end{thebibliography}
\end{document}